\magnification=1200
\baselineskip=14pt
\def\p{\wp^{(0)}}
\def\N{{\cal N}}
\def\Tr{{\rm Tr}}
\def\Det{{\rm Det}}

\rightline{UCLA/98/TEP/23}
\rightline{Columbia/Math/98}

\bigskip
\bigskip
\bigskip

\centerline{{\bf ORDER PARAMETERS, FREE FERMIONS, AND
CONSERVATION LAWS}}

\medskip

\centerline{{\bf FOR CALOGERO-MOSER SYSTEMS}
\footnote*{Research supported in part by
the National Science Foundation under grants PHY\-95-31023 and DMS-98-00783.}}

\bigskip
\bigskip
\bigskip

\centerline{{\bf Eric D'Hoker}${}^1$ 
            {\bf and D.H. Phong} ${}^2$}
\bigskip
\centerline{${}^1$ Department of Physics}
\centerline{University of California, Los Angeles, CA 90024, USA}
\centerline{Institute for Theoretical Physics}
\centerline{University of California, Santa Barbara, CA 93106, USA}
\bigskip
\centerline{${}^2$ Department of Mathematics}
\centerline{Columbia University, New York, NY 10027, USA}

\bigskip
\bigskip
\bigskip

\centerline{\bf ABSTRACT}

\bigskip

The classical order parameters for the $\N=2$ supersymmetric $SU(N)$ gauge
theory  with matter in the adjoint representation are exhibited explicitly as
conservation laws for the elliptic Calogero-Moser system.
Central to the construction are certain elliptic function identities, which
arise from considering Feynman diagrams in a theory of free fermions with
twisted boundary conditions.

\vfill\break

\centerline{\bf I. INTRODUCTION}

\bigskip

It has been known for a long time that the elliptic Calogero-Moser system
$$
p_i=\dot x_i,
\qquad
\dot p_i=m^2\sum_{j\not=i}\wp'(x_i-x_j),
\qquad \quad 1\leq i,j\leq N,
\eqno(1.1)
$$
is completely integrable, in the sense that it admits a Lax pair of
operators $L(z)$, $M(z)$ with a spectral parameter $z$ [1]. Here $\wp(z)$ is the
Weierstrass $\wp$-function on a fixed torus 
$\Sigma = {\bf C}/(2\omega_1{\bf Z}+2\omega_2{\bf Z})$ of modulus
$\tau=\omega_2/\omega_1$. The spectral curves
$$
\Gamma=\{(k,z);\  \det(kI-L(z))=0\},
\eqno(1.2)
$$
form an $N$-dimensional family of branched covers of the torus $\Sigma$.
More recently, in connection with Seiberg-Witten solutions of
four-dimensional $SU(N)$ supersymmetric gauge theories [2-5],
we have found that the spectral curves (1.2) admit a
natural parametrization of the form [5]
$$
\det\big(\lambda I-L(z)\big) =
{\vartheta_1({1\over 2\omega_1}(z-m{\partial\over\partial k})|\tau)
\over
\vartheta _1 ({z\over 2\omega_1}|\tau)}H(k)\bigg\vert_{k=\lambda+mh_1(z)}
\eqno(1.3)
$$
where $H(k)\equiv\prod_{i=1}^N(k-k_i)$ is a monic polynomial
of degree $N$, and the shift $h_1(z)$ is given by 
$h_1(z)=\partial_z\,\log\,\vartheta_1({z\over 2\omega_1}|\tau)$. 
From the point of view of four-dimensional
gauge theories, the zeroes $k_i$ of $H(k)$ have a very compelling
interpretation: they are 
the classical order parameters of the theory
(c.f. (1.5) in [5]). From the point of view of
Calogero-Moser systems, they are by construction integrals of motion
of the system. However, the derivation of $H(k)$ in [5]
did not provide explicit expressions for the $k_i$'s in terms
of the Calogero-Moser dynamical variables $(x_i,p_i)$. The goal
of the present paper is to solve this problem. In the process,
we also find an intriguing link between Calogero-Moser
systems and free fermions on a torus $\Sigma$.

\bigskip

To state the main result, we require the following notation. Let
$\sigma_m(k_1,\cdots,k_N)=\sigma_m(k)$ be the $m$-symmetric function of the 
$k_i$'s, as in
$$
H(k)=\prod_{i=1}^N(k-k_i)=\sum_{m=0}^N(-)^{m}\sigma_{m}(k_1,\cdots,k_N)\
k^{N-m},
\eqno(1.4)
$$
Let $\sigma_m(p)$ be the $m$-symmetric function of the momenta $p_i$'s,
$1\leq i\leq N$. If $S$ is any subset
of $\{1,2,\cdots,N\}$ consisting of $|S|$ numbers, we denote by
$\sigma_m(p_S)$ the $m$-symmetric function of $p_S=\{p_i; i\in S\}$,
for any integer $m$ with $m\leq |S|$. The complement of $S$ in $\{1,2,\cdots\}$
is denoted by $S^*$. If $S$ consists of only two elements $\{i,j\}$,
and $f(x)$ is an even function, we shall often write $f(S)$
for $f(x_i-x_j)$. All subsets $S$, $p_S$ are unordered, unless stated 
explicitly otherwise. Finally, it is convenient to introduce the following
modification of the Weierstrass $\wp$-function
$$
\p(z)= 
\wp(z)+{\eta_1\over\omega_1}.
\eqno(1.5)
$$
Here, $\eta _1$ and $\eta _2$ are the periods dual to $\omega_1$ and $\omega_2$.
Observe that $\wp^{(0)}(z)\rightarrow 0$ as $q\rightarrow 0$ and 
$z\rightarrow\infty$.
Then

\bigskip

\noindent
{\bf Main Theorem}. {\it The order parameters $k_i$, $1\leq i\leq N$, of the
gauge theory are related to the Calogero-Moser dynamical variables
$(x_i,p_i)$ by the following relations. For any integer $K$ with
$0\leq K\leq N$, we have}
$$
\sigma_K(k)
=\sigma_K(p)
+\sum_{l=1}^{[K/2]}m^{2l}
\sum_{|S_i\cap S_j|=2\delta_{ij}\atop 1\leq i,j\leq l}
\sigma_{K-2l}\big(p_{(\cup_{i=1}^lS_i)^{*}}\big)
\prod_{i=1}^l\p(S_i).
\eqno(1.6)
$$

\bigskip

As mentioned earlier, this theorem is partly motivated by current
investigation of $\N=2$ supersymmetric four-dimensional gauge theories [6-7]. 
The Wilson effective action of such theories is dictated by the spectral
curves of integrable models (see e.g. [8-10] for reviews). 
But it is still unclear whether
the dynamical variables of the integrable models have any direct
interpretation in the context of gauge theories. The preceding theorem
can be viewed as a step in addressing this question.

\medskip
In another direction, spectral curves have recently been obtained for
elliptic Calogero-Moser systems defined by general Lie algebras ${\cal G}$ 
[11-13]
and supersymmetric ${\cal G}$ gauge theories with matter in the adjoint
representation [11-15]. However, except in the case of $D_n$ [13] (see also
[15]),
a convenient parametrization such as (1.3) is still not available.
Such a parametrization is for example particularly
valuable in evaluating instanton corrections
to the prepotential [5].
It is conceivable that a deeper understanding of the order parameters
$k_i$ in the above $SU(N)$ case, as well
as the elliptic function identities found
in the present paper, may shed light on this issue.

\medskip
Finally, we mention some related problems in the theory of integrable
models proper. The symplectic
structure of Calogero-Moser systems is attracting considerable 
attention
[16-17]. The integrals of motion (1.6)
may be relevant to the well known problem of constructing $R$-matrices
for Calogero-Moser systems (c.f. [18-19]). They may also
be of interest in the rational and trigonometric cases [20].
In particular, in the trigonometric case,
the gauge order parameters $k_i$
have been
useful in the study of Toeplitz determinants, symplectic volumes, 
and thermodynamic limits [21].

\vfill\break

\centerline{\bf II. MAIN IDENTITIES AND PROOF OF THE THEOREM}

\bigskip

We divide the proof of the Main Theorem into several steps.

\bigskip

\noindent
(I) In the first step, the defining identity (1.3) for the integrals $k_i$ is
rewritten in terms of determinants $D(S)$ similar to $\det\, L(z)$, but with all
diagonal entries set to 0. More precisely, recall that the Lax pair $L(z)$,
$M(z)$ for the elliptic Calogero-Moser system is given by [1]
$$
\eqalignno{
L_{ij}(z)&=p_i\delta_{ij}-m(1-\delta_{ij})\Phi(x_i-x_j,z),\cr
M_{ij}(z)&=m\delta_{ij}\sum_{k\not=i}\wp(x_i-x_k)
+m(1-\delta_{ij})\Phi'(x_i-x_j,z) 
&(2.1)\cr}
$$
with 
$$
\Phi(x,z)={\sigma(z-x)\over\sigma(z)\sigma(x)}e^{x\zeta(z)},
\eqno(2.2)
$$
Here $\sigma(z)$, $\zeta(z)$ are the usual Weierstrass elliptic functions
(c.f. Erdelyi [22]). Let $S=\{\alpha(1),\alpha(2),\cdots,\alpha(K)\}$ be a
subset of $\{1,2,\cdots,N\}$ with $|S|=K$ (distinct) elements.  We define $D(S)$
to be the following $K\times K$ determinant 
$$
D(S)=\det \bigg [(1-\delta_{ij})(\Phi(x_{\alpha(i)}-x_{\alpha(j)},z)\bigg ]
\eqno(2.3)
$$
There is no ambiguity in this definition since the 
right hand side of (2.3) is independent of the ordering of $S$. 
The identity (1.3)
is then equivalent to
$$
\sum_{p+q=N-n}\pmatrix{n+q\cr n\cr}m^qA_q(-)^p\sigma_p(k)
=
\sum_{|S|+l=N-n}(-)^l\sigma_l(p_{S^*})m^{|S|}D(S),
\ 1\leq n\leq N,
\eqno(2.4)
$$
In these identities the summation on the right hand side 
is over all subsets $S$ of $\{1,2,\cdots,N\}$ and the functions $A_K(z)$
are given explicitly in terms of $\vartheta$-functions
$$
\eqalignno{
A_K(z)&=\sum_{n=0}^K\pmatrix{K\cr n\cr}(-)^nh_n(z)h_1(z)^{K-n}\cr
h_n(z)&={\partial_z^n\,\vartheta_1({z\over 2\omega_1}|\tau)
\over \vartheta_1({z\over 2\omega_1}|\tau)}
&
(2.5)\cr}
$$ 
To establish (2.4), we note that
$$
{\vartheta_1 ({1\over 2\omega_1}(z-m{\partial\over\partial k})|\tau)
\over
\vartheta _1 ({z\over 2\omega_1}|\tau)}H(k)
=\sum_{n=0}^N{h_n(z)\over n!}(-m)^nH^{(n)}(k)
\eqno(2.6)
$$
Substituting in $k=\lambda+mh_1(z)$ and expanding at $\lambda$, we find
$$
\eqalignno{
{\vartheta_1({1\over 2\omega_1}(z-m{\partial\over\partial k})|\tau)
\over
\vartheta _1 ({z\over 2\omega_1}|\tau)}H(k)\bigg\vert_{k=\lambda+mh_1(z)}
&=\sum_{n=0}^N\sum_{q=o}^{N-n}{h_n(z)h_1(z)^q\over n!q!}(-)^nm^{n+q}H^{(n+q)}
(\lambda)\cr
&=\sum_{K=0}^NA_K(z){m^KH^{(K)}(\lambda)\over K!}
&(2.7)\cr}
$$
In terms of $\sigma_p(k)$, the polynomials $H^{(K)}(\lambda)$
are just given by
$$
H^{(K)}(\lambda)=\sum_{p=0}^{N-K}(-)^m{(N-p)!\over (N-K-p)!}\sigma_p(k)
\lambda^{N-K-p}
\eqno(2.8)
$$ 
This implies that the right hand side of (1.3) can be rewritten as
$$
{\vartheta_1({1\over 2\omega_1}(z-m{\partial\over\partial k})|\tau)
\over
\vartheta _1 ({z\over 2\omega_1}|\tau)}H(k)\bigg\vert_{k=\lambda+mh_1(z)}
=
\sum_{n=0}^N\lambda^n\sum_{p+q=N-n}\pmatrix{n+q\cr n\cr}
A_q(z)m^q(-)^p\sigma_p(k).
\eqno(2.9)
$$
On the other hand, the determinant $\det(\lambda I-L(z))$ on the left
hand side of (1.3) can be expanded as
$$
\det(\lambda I-L(z))
=\sum_{K=0}^Nm^K
\sum_{|S|=K}\big[\prod_{i\in S^*}(\lambda-p_i)\big]D(S)
\eqno(2.10)
$$
Evidently,
$$
\prod_{i\in S^*}(\lambda-p_i)
=\sum_{l=0}^{N-K}(-)^l\sigma_l(p_{S^*})\lambda ^{N-K-l}
$$ 
so that the left hand side of (1.3) becomes
$$
det(\lambda I-L(z))
=\sum_{n=0}^N\lambda^n
\sum_{l=0}^{N-n}m^{N-l-n}\sum_{|S|=N-l-n}(-)^l\sigma_l(p_{S^*})D(S).
\eqno(2.11)
$$
Comparing (2.9) with (2.11) gives the desired identities (2.4).

\bigskip

\noindent
(II) The second step in the proof of the main theorem
consists of, in a sense, separating in the determinant $D(S)$
the dependence on the insertion points $x_{\sigma(j)}\in S$ 
from the dependence on the spectral parameter $z$.
More precisely, we can write 
$$
D(S)=\sum_{l=0}^{[K/2]}B_{K-2l}(z)
\sum_{|S_i\cap S_j|=2\delta_{ij}\atop 1\leq i,j\leq l}
\prod_{i=1}^l\p({S_i})
\eqno(2.12)
$$
where $\p(S_i)=\p(x_a-x_b)$ if $S_i=\{a,b\}$
is the convention introduced in Section I.
To describe the coefficients $B_M(z)$, it is convenient to
introduce the notation
$$
\wp^{(n)}(z)=\big({\partial\over\partial z}\big)^n\p(z), \qquad n\in {\bf N}.
\eqno(2.13)
$$
Then the coefficients $B_K(z)$ can be expressed as
$$
B_K(z)
=(-)^K\sum_{K=2L_2+3L_3+\cdots}(-)^{\sum_{n=2}^{\infty}L_n}
{K!\over \prod_{n=2}^{\infty}(L_n!)\prod_{n=2}^{\infty} (n!)^{L_n}}
\prod_{n=2}^{\infty}\big[\wp^{(n-2)}(z)\big]^{L_n}
\eqno(2.14)
$$
The proof of the identity (2.14) is the lengthiest part of our
argument, and we postpone it until Section III.

\bigskip

\noindent
(III) The third step in the proof of the Main Theorem is to show that
the coefficients $A_K(z)$ and $B_K(z)$ are actually equal
$$
A_K(z)=B_K(z),\ \ K=1,2,3,\cdots.
\eqno(2.15)
$$
Evidently, $A_1(z)=B_1(z)=0$, while $A_2(z)=h_2(z)-h_1(z)^2$ and 
$B_2(z)=-\p(z)$. Since the Weierstrass $\sigma$-function and the
Jacobi $\vartheta$-function are related by
$$
\eqalignno{
\sigma(z)&=2\omega_1 e^{{\eta_1\over 2\omega_1}z^2}
{\vartheta_1({z\over 2\omega_1}|\tau)\over\vartheta_1'(0|\tau)}\cr
\wp(z)&=-\partial_z^2\,log\,\sigma(z)
=-{\eta_1\over\omega_1}-\partial_z^2\,\log \vartheta_1({z\over 2\omega_1}|\tau)
&(2.16)
\cr}
$$
we also have $A_2(z)=B_2(z)$. It suffices then to show that both
the $A_K(z)$'s and the $B_K(z)$'s obey the same two-step recursive relation
$$
\eqalignno{
A_{K+1}(z)&=-A_K'(z)-K\p(z)A_{K-1}\cr
B_{K+1}(z)&=-B_K'(z)-K\p(z)B_{K-1}&(2.17)\cr}
$$
for $K\geq 2$. The recursive relation for $A_{K+1}(z)$ is easily
established by differentiating $A_K(z)$, and using the fact that
$$
\eqalign{
h_n'(z) & =-h_n(z)h_1(z)+h_{n+1}(z), \cr
h_1'(z) & = -h_1(z)^2+h_2(z)=-\p(z).\cr}
\eqno(2.18)
$$
To establish the recursive relation for $B_K(z)$,
we define $B_0(z)$ to be $1$,
introduce an additional variable $y$, and consider the generating
function
$$
\sum_{K=0}^{\infty}{B_K(z)\over K!}y^K
={\rm exp}\,\bigg [\sum_{n=2}^{\infty}{(-)^{n+1}\over n!}\wp^{(n-2)}(z)y^n
\bigg ].
\eqno(2.19)
$$
Differentiating with respect to $y$ gives a recursive relation
for $B_{K+1}(z)$
$$
B_{K+1}(z)
=\sum_{n=1}^K\pmatrix{K\cr n\cr}(-)^{n}\wp^{(n-1)}(z)B_{K-n}(z),
\eqno(2.20)
$$
while differentiating with respect to $z$ gives a recursive
relation for $B_K'(z)$
$$
B_K'(z)=\sum_{n=2}^K\pmatrix{K\cr n\cr}(-)^{n+1}\wp^{(n-1)}(z)B_{K-n}(z).
\eqno(2.21)
$$
Comparing (2.20) with (2.21) gives the desired recursive
relation (2.17). The identity $A_K(z)=B_K(z)$ is established.

\bigskip

\noindent
(IV) With the identities (I-III), we can now prove the theorem.
We fix integers $N$ and $n$, with $n\leq N$. The sum in the
right hand side of (2.4) is over all subsets $S$ of $\{1,2,\cdots,N\}$.
Setting $|S|=K$ and using (II) and (III),
we can express it as
$$
\sum_{K=0}^{N-n}(-)^{N-n-K}m^K
\sum_{S,|S|=K}\sigma_{N-n-K}(p_{S^*})
\sum_{q+2j=K}A_q(z)\sum_{|S_i\cap S_j|=2\delta_{ij}
\atop S_i\subset S}\p(S_1)\cdots\p(S_j)
\eqno(2.22)
$$
Introduce the index $p=N-n-q$ (not to be confused with the
Calogero-Moser momenta!). Then the order of summations in (2.22)
can be interchanged to produce
$$
\sum_{q=0}^{N-n}
m^qA_q(z)(-)^p\sum_{j=0}^{[p/2]}
\sum_{S, |S|=q+2j}
\sigma_{p-2j}(p_{S^*})m^{2j}
\sum_{|S_i\cap S_j|=2\delta_{ij}
\atop S_i\subset S}\p(S_1)\cdots\p(S_j).
\eqno(2.23)
$$
However, for each $j$, we have the following combinatorial
identity
$$
\eqalignno{\sum_{S, |S|=q+2j}
&\sigma_{p-2j}(p_{S^*})\, m^{2j}
\sum_{|S_i\cap S_j|=2\delta_{ij}
\atop S_i\subset S}\p(S_1)\cdots\p(S_j)
\cr
&=\pmatrix{n+q\cr n\cr}
\sum_{|S_i\cap S_j|=2\delta_{ij}}
\sigma_{p-2j}(p_{(\cup_{i=1}^jS_i)^*})
\p(S_1)\cdots\p(S_j).
&(2.24)\cr}
$$
for $p-2j\geq 0$. In fact, by permutation invariance, the expressions
on the two sides of (2.24) are proportional. To determine
the coefficient of proportionality, we compare the coefficients
of the term $\p(x_1-x_2)\cdots\p(x_{2j-1}-x_{2j})p_{2j+1}\cdots p_p$
which occurs on both sides.
In the sum on the right hand side of (2.24), such a term
occurs exactly once. On the other hand, such a term occurs
in the sum on the left hand side whenever we can choose a subset
$S$ of size $q+2j$, containing
$\{1,\cdots,{2j}\}$, and not containing
$\{{2j+1},\cdots,p\}$. This means that $S$ consists of $\{1,\cdots,2j\}$,
together with $q$ more elements in $\{p+1,\cdots,N\}$. There are
exactly
$$
\pmatrix{N-p\cr q\cr}=\pmatrix{n+q\cr q\cr}=\pmatrix{n+q\cr n\cr}
\eqno(2.25)
$$
such choices. Thus the expression (2.23) becomes
$$
\sum_{q=0}^{N-n}
m^qA_q(z)(-)^p\pmatrix{n+q\cr n}
\sum_{|S_i\cap S_j|=2\delta_{ij}}
\sigma_{p-2j}(p_{(\cup_{i=1}^jS_i)^*})
\p(S_1)\cdots\p(S_j),
\eqno(2.26)
$$
with $p=N-n-q$. Comparing with the left hand side
of (2.4) gives the theorem. 

\bigskip
\bigskip

\centerline{\bf III. THE DETERMINANT $D(S)$ AND FREE FERMIONS}

\bigskip

It remains to establish the identities in (II) of Section II
for the determinants $D(S)$. For this we need the notion of ``$k$-cycle",
which we now describe. Let $\{1,2,\cdots,k\}$ be any set of $k$ indices,
which we chose to be the first $k$ integers just for notational
convenience. Then the expression
$$
\eqalign{
\Phi_{12}\Phi_{23} & \cdots\Phi_{(k-1)k}\Phi_{k1} \cr
\equiv \ &\Phi(x_1-x_2,z)\Phi(x_2-x_3,z)
\cdots\Phi(x_{k-1}-x_k,z)\Phi(x_k-x_1,z) \cr}
\eqno(3.1)
$$
is a single-valued, meromorphic function of all insertion points
$x_1,\cdots,x_k$, as well as of the spectral parameter $z$. 
Here we have made use of the monodromy properties of the function
$\Phi(x,z)$ as a function of $x$
$$
\eqalignno{
\Phi(x+2\omega_a,z)=&\Phi(x,z)e^{2\omega_a\zeta(z)-2\eta_a z}\cr
{1\over 2\pi}\partial_{\bar x}\Phi(x-y,z)=&\delta(x-y)
&(3.2)\cr}
$$
(As a function of $z$, $\Phi(x,z)$ is already by itself
single-valued on the torus $\Sigma$.)
It is useful to note that in
expressions such as (3.1), the function $\Phi(x,z)$ can be effectively
replaced by $\sigma(z-x)/\sigma(z)\sigma(x)$. 
We define a $k$-cycle to be the sum of all inequivalent
expressions (3.1) under permutations of the indices $1,2,\cdots,k$. 
Since (3.1)
is evidently invariant under shifts in the indices $1,2,\cdots,k$,
this sum corresponds to a sum over ${\bf S}_k/{\bf Z}_k$, where
${\bf S}_k$ is the group of permutations of $k$ elements. 
Equivalently, we can fix an index, say $k$, and write a $k$-cycle
as
$$
C_k(x_1,\cdots,x_k;z)=\sum_{\alpha\in{\bf S}_{k-1}}
\Phi_{k\alpha(1)}\Phi_{\alpha(1)\alpha(2)}
\cdots\Phi_{\alpha(k-1)k},
\eqno(3.3)
$$
identifying in effect ${\bf S}_k/{\bf Z}_k$ with the group
${\bf S}_{k-1}$ of permutations of $k-1$ elements.
It is easy to verify that, for $k\geq 3$,
the $k$-cycle $C_k(x_1,\cdots,x_k;z)$ is actually
a function $C_k(z)$ independent
of the insertion points $\{x_1,x_2,\cdots,x_k\}$.
Indeed, viewed as a function of say $x_1$, it is meromorphic
and has simple poles at the other insertion points
$x_2,\cdots, x_k$. The residues at each of these poles however
cancel out between the various terms in (3.2), so that
the $k$-cycle is actually constant in each $x_j$.
The main problem is then to determine the dependence on $z$
of $k$-cycles. The identity central to our approach is the
following
$$
C_k(z)=\sum_{\alpha\in{\bf S}_{k-1}}\Phi_{k\alpha(1)}\Phi_{\alpha(1)\alpha(2)}
\cdots\Phi_{\alpha(k-1)k}
=\wp^{(k-2)}(z),\qquad k=3,4,\cdots.
\eqno(3.4)
$$
(For $k=2$ the 2-cycle is not independent of the insertion points.
In fact, we have
$$
\Phi_{12}\Phi_{21}=\wp(z)-\wp(x_1-x_2)
=\p(z)-\p(x_1-x_2),
\eqno(3.5)
$$
a well-known and basic identity in the
theory of elliptic Calogero-Moser systems.) Postponing for the moment
the proof of (3.4), we return to the study of the determinants $D(S)$.

\bigskip

\noindent
{\it Exact Formulas for $D(S)$}

Since the diagonal elements of the matrix $D(S)$ all vanish, the determinant
can be expanded as
$$
D(S)
=\sum_{\alpha}(-)^{\alpha}\Phi_{1\alpha(1)}\Phi_{2\alpha(2)}
\cdots\Phi_{K\alpha(K)}
\eqno(3.6)
$$
where the summation is only over permutations $\alpha$
without any fixed point. But it is readily seen that
any permutation $\alpha$ 
without fixed point corresponds to a decomposition
of the index set $\{1,2,\cdots,K\}$ into disjoint subsets $S_j$
of at least two elements,
in each of which $\alpha$ acts as a shift. Since the sign of a shift
on $N$ elements is $(-)^{N+1}$, the sign of $\alpha$
is $(-)^{K+l}$, where $l$ is the number of subsets $S_j$.
All permutations
without fixed points can be generated this way, by following up
the decomposition of the index set into smaller sets $S_j$
with permutations within each smaller set $S_j$.
Taking these ``internal" permutations into account, the contribution
of each decomposition $S=\cup_{j=1}^l S_j$ to the determinant (3.6)
is a product of $k$-cycles
$$
\prod_{j=1}^l\big[|S_j|-{\rm cycles}\big]\eqno(3.7)
$$
In view of (3.4) and (3.5),
this establishes the fact that the determinant $D(S)$
must be of the form (2.12), for some as yet complicated
coefficients $B_{K-2l}(z)$, $1\leq l\leq [K/2]$.

\medskip

To determine $D(S)$, it suffices to determine the ``constant term" $B_K(z)$.
This is because identities of the form (2.12) will be
established inductively, by examining the poles of both
sides of the equation in each of the variables $x_i$.
For example, consider the double pole in the variable $x_1$, near
the value $x_2$. For the left hand side, it is
$$
-\Phi_{12}\Phi_{21}\times D(S\setminus\{1,2\})
=[\p(x_1-x_2)-\p(z)]\times D(S\setminus\{1,2\})
\eqno(3.8)
$$
For the right hand side, it is
$$
\bigg[\sum_{l=1}^{[K/2]}B_{K-2l}(z)\sum_{|S_i\cap S_j|=2\delta_{ij}
\atop S_i\subset S\setminus\{1,2\}}
\p(S_2)\cdots\p(S_l)\bigg]\p(x_1-x_2).
\eqno(3.9)
$$
By induction, the expression between brackets
is indeed $D(S\setminus\{1,2\})$.
Similarly, the simple poles cancel.
This shows that $D(S)$ is determined up
to an additive function of $z$ only.

\medskip

We can derive now the explicit formula (2.14) for $B_K(z)$.
Since we are restricting our attention to the constant term $B_K(z)$
in the expansion (3.6-3.7) for $D(S)$, we can replace even the 2-cycles
in (3.7) by $\p(z)$. With this simplification, the contribution of (3.7)
to $B_K(z)$ is just
$$
\prod_{j=1}^l\wp^{(|S_j|-2)}(z)
\eqno(3.10)
$$
Now the exact subsets $S_j$ themselves no longer matter, and the only
relevant information is their size $|S_j|$. For each partition of $K$
into $L_2$ subsets of $2$ elements, $L_3$ subsets of $3$ elements,
etc.
$$
K=2L_2+3L_3+4L_4+\cdots=\sum_{n=2}^{\infty}nL_n
\eqno(3.11)
$$
the expression (3.10) becomes
$$
\prod_{n=2}^{\infty}\big[\wp^{(n-2)}(z)\big]^{L_n}
\eqno(3.12)
$$
Now the number of ways of selecting $L$ (unordered) sets of $n$ elements each
from an ensemble of $N$ elements is 
$$
{1\over L!}{N!\over (n!)^L (N-nL)!}.
\eqno(3.13)
$$
Thus the total number of terms of the form (3.12) is
$$
\eqalignno{{1\over L_2!}{N!\over (2!)^{L_2} (N-2L_2)!}
\times
{1\over L_3!}&{(N-2L_2)!\over (3!)^{L_3} (N-2L_2-3L_3)!}
\times
{1\over L_4!}{(N-2L_2-3L_3)!\over (4!)^{L_4} (N-2L_2-3L_3-4L_4)!}
\times\cr
&\cdots
={N!\over\prod_{n=2}^{\infty}L_n!\prod_{n=2}^{\infty}(n!)^{L_n}}
&(3.14)\cr}
$$
Altogether, this establishes the formula (2.14).

\bigskip

\noindent
{\it Free Fermions and $k$-Cycles as Feynman Diagrams}

Finally, we turn to the proof of the fundamental identity (3.4).
The main idea is to view $k$-cycles $C_k(x_1,\cdots,x_k;z)$
as the one-loop amplitude
in a theory of free fermions with twisted boundary conditions
on a torus and fermion propagator $\Phi(x-y,z)$. Here
$z$ is viewed as a fixed parameter. 

First, since $C_k(x_1,\cdots,x_k;z)$
is independent of $x_i$ anyway, we may just as well integrate
each $x_i$ over the torus $\Sigma$ with area $\tau_2$,
$$
\eqalignno{
C_k(z)
&=
\int _\Sigma {d^2x_1\over\tau_2} \cdots \int _\Sigma {d^2x_k\over\tau_2}
C_k(x_1,\cdots,x_k;z)\cr
&=
(k-1)!\int _\Sigma {d^2x_1\over\tau_2} \cdots \int _\Sigma {d^2x_k\over\tau_2}
\Phi(x_1-x_2,z)\cdots\Phi(x_k-x_1,z)
&(3.15)\cr}
$$
where the factor $(k-1)!$ comes out since integration in each variable
automatically takes care of symmetrization. In this integrated
form, the one-loop amplitude $C_k(z)$ has an even simpler
interpretation, which we now develop. Starting from the free massless fermion
propagator, 
$$
 \Phi (x-y,z),
\eqno(3.16)
$$
we may construct the ``full" propagator of a fermion in the presence of a
constant (background) gauge potential with strength $m$, by summing up the
effects of repeated gauge potential coupling operator insertions. The ``full"
fermion propagator may thus be defined by the geometric series
$$
S(x-y|z,m)
=\Phi(x-y,z)+{m\over\tau_2}\int _\Sigma d^2y_1\Phi(x-y_1,z)\Phi(y_1-y,z)
+\cdots
\eqno(3.17)
$$
or in terms of the recursive relation
$$  
S(x-y|z,m)
=\Phi(x-y,z)+{m\over\tau_2}\int _\Sigma d^2y_1\Phi(x-y_1,z) S(y_1-y|z,m).
\eqno(3.18)
$$
The $k$-cycles $C_k(z)$ are now easily gotten as the $k-1$ derivatives
with respect to $m$ of the propagator $S(x-y|z,m)$ at coincident points
$S(0|z,m)$, as we shall use below in (3.26).

Equivalently, we may characterize $S(x-y|z,m)$ by its monodromy
and differential equation, which follow from the analogous
properties of the propagator $\Phi(x-y,z)$, and the definitions (3.17) and
(3.18),
$$
\eqalignno{
S(x+2\omega_a|z,m)=& \ S(x|z,m)e^{2\omega_a\zeta(z)-2\eta_a z}\cr
\bar D S(x-y|z,m)=&\ \delta(x-y)
&(3.19)\cr}
$$
where we introduce the $\bar D$ and $D$ operators by
$$
\bar D={1\over 2\pi}\partial_{\bar x}-{m\over\tau_2},
\qquad
D={1\over 2\pi}\partial_{ x}-{\bar m\over\tau_2}.
\eqno(3.20)
$$
These operators are precisely the Dirac operators on the torus $\Sigma$
in the presence of a constant gauge potential with strength $m$ for left-
and right-movers respectively.

Now we need $C_k(z)$, gotten by one closed loop, i.e. by $S(0|z,m)$. Thus it
suffices to evaluate the determinant, since
$$
{\partial\over\partial m}\log\, \Det \bar D D
={\partial\over\partial m}\Tr \,\log\, \bar D
=-{1\over\tau_2} \Tr\bar D ^{-1} = -S(0|z,m)
\eqno(3.21)
$$
The eigenvalues of $\bar D$ on the space of functions
with monodromy as in (3.19) can be determined as usual.
They are given by
$$
\lambda_{n_1n_2}
={1\over\tau_2}(n_1\tau-n_2+m+z), \qquad n_1,n_2\in{\bf Z}
\eqno(3.22)
$$
We recall from [23], p. 1002, that the determinant
of a Dirac operator $\bar D$ is given by
$$
\Det\bar D={\vartheta_{\nu_1\nu_2}(0|\tau)\over\eta(\tau)},
\eqno(3.23)
$$
if its eigenvalues are of the form
$$
\lambda_{n_1n_2}={1\over\tau_2}
\big[(n_1+{1\over 2}-{1\over 2}\nu_1)\tau
-(n_2+{1\over 2}-{1\over 2}\nu_2)\big].
\eqno(3.24)
$$
In the present case, $\nu_1=1$, $\nu_2=1-2(m+z)$, and we obtain
$$
\Det\bar D={\vartheta_1(z+m|\tau)\over\eta(\tau)}
\eqno(3.25)
$$
Returning to the $k$-cycles $C_k(z)$, we can write
$$
C_k(z)=({\partial\over\partial m})^{k-1}S(0|z,m)\bigg\vert_{m=0}
=-({\partial\over\partial m})^k\,log\,{\vartheta_1(z+m|\tau)\over\eta(\tau)}
\bigg|_{m=0}
\eqno(3.26)
$$
The desired identity (3.4) follows now
from the elliptic function identity
(2.16). The proof of the main theorem is complete.

\vfill\break

\centerline{\bf REFERENCES}

\bigskip

\item{[1]} Krichever, I., ``Elliptic solutions of the Kadomtsev-Petviashvili
equation and integrable systems of particles",
Funct. Anal. Appl. {\bf 14} (1980) 282-290.

\item{[2]} Donagi, R. and E. Witten, ``Supersymmetric Yang-Mills theory
and integrable systems", Nucl. Phys. {\bf B 460} (1996) 299-334,
hep-th/9510101.

\item{[3]} Martinec, E. and N. Warner, ``Integrable models and
supersymmetric gauge theory", Nucl. Phys. {\bf B 459} (1996) 97-112,
hep-th/9609161.

\item{[4]} Martinec, E., ``Integrable structures in supersymmetric
gauge and string theory", hep-th/9510204.

\item{[5]} D'Hoker, E. and D.H. Phong,
``Calogero-Moser systems in SU($N$) Seiberg-Witten
theory", hep-th/9709053, Nucl. Phys. {\bf B 513} (1998), 405.

\item{[6]} Seiberg, N. and E. Witten, ``Electro-magnetic duality,
monopole condensation, and confinement in N=2 supersymmetric
Yang-Mills theory", Nucl. Phys. {\bf B 426} (1994), 19,
hep-th/9407087.

\item{[7]} Seiberg, N. and E. Witten, ``Monopoles, duality, and
chiral symmetry breaking in N=2 supersymmetric QCD",
Nucl. Phys. {\bf B 431} (1994) hep-th/9410167.

\item{[8]} Lerche, W., ``Introduction to Seiberg-Witten theory and its
stringy origins", Proceedings of the {\it Spring School and Workshop
in String Theory}, ICTP, Trieste (1996), hep-th/9611190.

\item{[9]} Donagi, R., ``Seiberg-Witten integrable systems",
alg-geom/9705010.

\item{[10]} Carroll, R., ``Prepotentials and Riemann surfaces",
hep-th/9802130.

\item{[11]} D'Hoker, E. and D.H. Phong, ``Calogero-Moser Lax pairs
with spectral parameters for general Lie algebras",
hep-th/9804124.

\item{[12]} D'Hoker, E. and D.H. Phong, ``Calogero-Moser and Toda systems
for twisted and untwisted affine Lie algebras",
hep-th/9804125.

\item{[13]} D'Hoker, E. and D.H. Phong, ``Spectral curves for super Yang-Mills
with adjoint hypermultiplet for general Lie algebras",
hep-th/9804126.

\item{[14]} Uranga, A.M., ``Towards mass-deformed N=4 SO(N) and Sp(K)
gauge theories from brane configurations", hep-th/9803054.

\item{[15]} Yokono, T., ``Orientifold four plane in brane
configurations and N=4 USp(2N) and SO(2N) theory",
hep-th/9803123.

\item{[16]} Gorsky, A. and N. Nekrasov,
``Elliptic Calogero-Moser systems from two-dimensional
current algebras", ITEP-NG/1-94, hep-th/9401021.

\item{[17]} Krichever, I.M. and D.H. Phong, ``Symplectic forms in
the theory of solitons", hep-th/9708170, to appear in
{\it Surveys in Differential Geometry, III}.

\item{[18]} Olshanetsky, M.A. and A. Perelomov, ``Classical integrable
finite-dimensional systems related to Lie algebras",
Phys. Reports {\bf 71} (1981) 313-400.

\item{[19]} Braden, H.W., ``R-matrices, Generalized Inverses,
and Calogero-Moser-Sutherland Models",
Proc. Conference on Calogero-Moser-Sutherland Models,
Montreal, 1997.

\item{[20]} Bordner, A.J., E. Corrigan, and R. Sasaki, ``Calogero-Moser
Models : A New Formulation", hep-th/9805106.

\item{[21]} Vaninsky, K.L., ``On explicit parametrization of spectral
curves for Moser-Calogero particles and its application",
June 1998 preprint, Kansas State University.

\item{[22]} Erdelyi, A., ed., {\it ``Higher Transcendental Functions"}, R.E. 
Krieger Publishing Co., Florida (1981).

\item{[23]} D'Hoker, E. and D.H. Phong, ``The geometry
of string perturbation theory", Rev. Modern Physics 80 (1988)
917-1065.

\end